\documentclass{vgtc}                          

\ifpdf
  \pdfoutput=1\relax                   
  \usepackage{graphicx}                
  \DeclareGraphicsExtensions{.pdf,.png,.jpg,.jpeg} 
\else
  \usepackage{graphicx}                
  \DeclareGraphicsExtensions{.eps}     
\fi%

\graphicspath{{figures/}{pictures/}{images/}{./}} 

\usepackage{microtype}                 
\PassOptionsToPackage{warn}{textcomp}  
\usepackage{textcomp}                  
\usepackage{mathptmx}                  
\usepackage{times}                     
\usepackage{cite}                      
\usepackage{tabu}                      
\usepackage{booktabs}                  

\usepackage{amsmath, amssymb}
\usepackage{type1cm}
\usepackage{enumitem}
\usepackage[whole]{bxcjkjatype}
\usepackage{balance}

\onlineid{0}

\vgtccategory{Research}

\vgtcinsertpkg

\DeclareMathAlphabet\mathbfcal{OMS}{cmsy}{b}{n}
\DeclareMathAlphabet\mathbfit{OML}{cmm}{b}{it}

\newcommand{\Mat}[1]{\mathbfit{#1}}
\newcommand{\Tensor}[1]{\mathbfcal{#1}}

\title{
Angular-based Edge Bundled Parallel Coordinates Plot \\
for the Visual Analysis of Large Ensemble Simulation Data}

\author{Keita Watanabe\thanks{e-mail: 225x226x@stu.kobe-u.ac.jp}\\ %
        \scriptsize Kobe University %
\and Naohisa Sakamoto\thanks{e-mail: naohisa.sakamoto@people.kobe-u.ac.jp}\\ %
     \scriptsize Kobe University %
\and Jorji Nonaka\thanks{e-mail: jorji@riken.jp}\\ %
     \scriptsize RIKEN R-CCS %
\and Yasumitsu Maejima\thanks{e-mail: yasumitsu.maejima@riken.jp}\\ %
     \scriptsize RIKEN R-CCS}

\teaser{
  \centering
  \includegraphics[width=\linewidth]{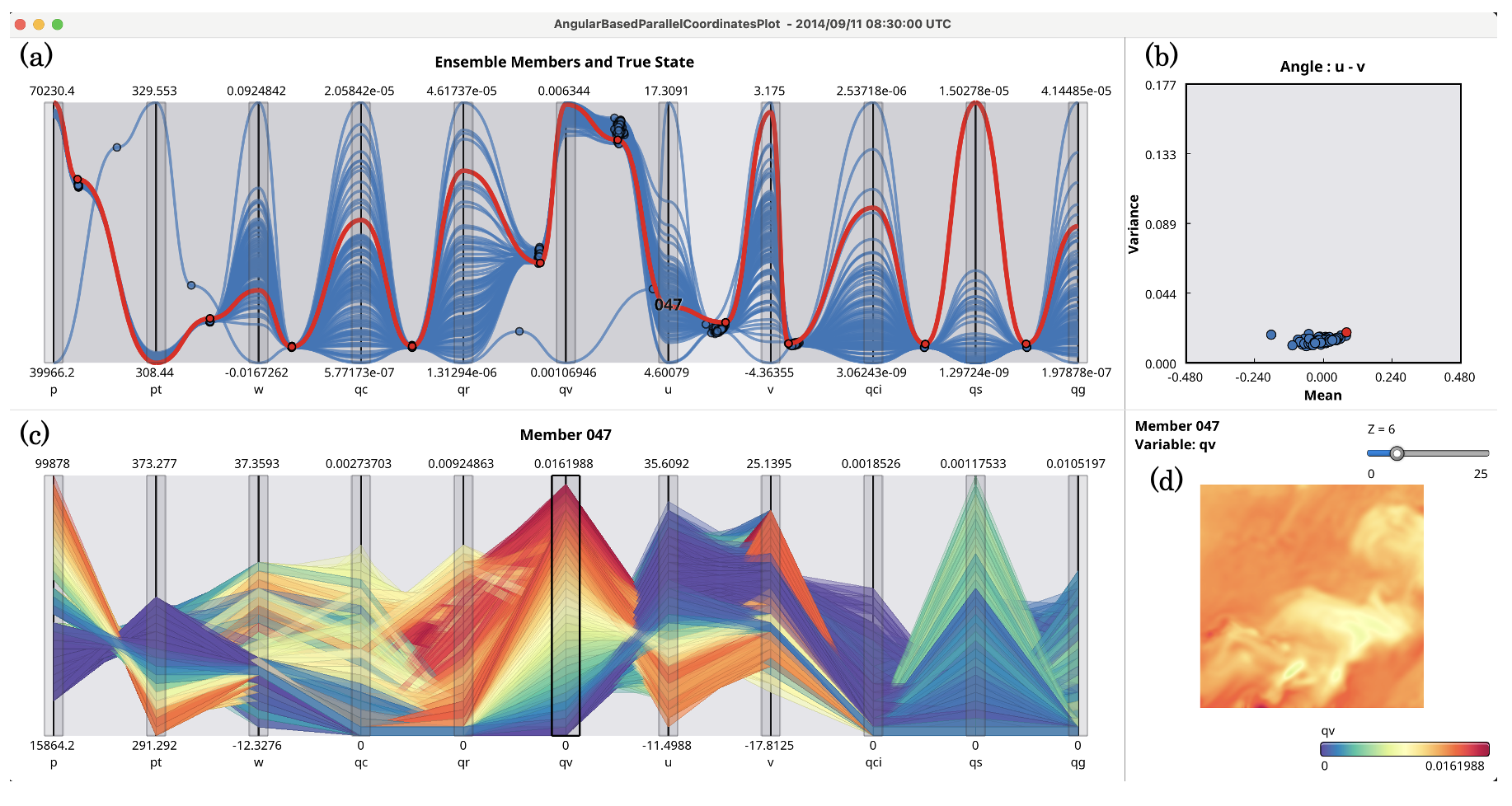}
  \caption{Overview of the implemented visual analytics system showing the analysis of a 100-member ensemble simulation data. It is composed of the following linked coordinated views: (a) Angular-based Parallel Coordinates Plot (APCP) view, (b) Angular Distribution Plot (ADP) view, (c) Binning Parallel Coordinates Plot (BPCP) view, and (d) Cross Sectional Plot (CSP) view.}
  \label{fig:teaser}
}

\abstract{With the continuous increase in the computational power and resources of modern high-performance computing (HPC) systems, large-scale ensemble simulations have become widely used in various fields of science and engineering, and especially in meteorological and climate science. It is widely known that the simulation outputs are large time-varying, multivariate, and multivalued datasets which pose a particular challenge to the visualization and analysis tasks. In this work, we focused on the widely used Parallel Coordinates Plot (PCP) to analyze the interrelations between different parameters, such as variables, among the members. However, PCP may suffer from visual cluttering and drawing performance with the increase on the data size to be analyzed, that is, the number of polylines. To overcome this problem, we present an extension to the PCP by adding B\'{e}zier curves connecting the angular distribution plots representing the mean and variance of the inclination of the line segments between parallel axes. The proposed Angular-based Parallel Coordinates Plot (APCP) is capable of presenting a simplified overview of the entire ensemble data set while maintaining the correlation information between the adjacent variables. To verify its effectiveness, we developed a visual analytics prototype system and evaluated by using a meteorological ensemble simulation output from the supercomputer Fugaku.
}

\CCScatlist{
  \CCScatTwelve{Human-centered computing}{Visu\-al\-iza\-tion}{Visu\-al\-iza\-tion techniques}{};
  \CCScatTwelve{Human-centered computing}{Visu\-al\-iza\-tion application domains}{Visual analytics}{}
}

\begin{document}

\firstsection{Introduction}

\maketitle




Extreme weather events have continuously caused worldwide human losses and widespread material and infrastructure damages~\cite{wmo2021extreme}. There is also a report that these disaster-class events are unfortunately increasing in frequency and intensity~\cite{ipcc2021extreme}. HPC-based ensemble numerical weather prediction (NWP) has been applied with the aim to increase the lead time of extreme weather event warnings~\cite{stensrud2009convective}~\cite{yano2018scientific}. Data assimilation which uses real-time observational data has become more commonly used in the ensemble simulations for periodically adjusting their initial conditions, and aiming for better predictions~\cite{tsuyuki2007}. In Japan, the use of HPC system and phased array weather radar (PAWR) has received increasing attention as a promising approach for tackling sudden localized torrential rains, also known as \textit{guerrilla rainstorms}, which have caused flooding damages and landslide disasters. For this purpose, flagship-class supercomputers (former K computer and current Fugaku) have been used for running large ensemble simulations with data assimilation. As an example, we can cite an undergoing ensemble simulation with more than 1,000 members on the supercomputer Fugaku~\cite{duc2021forecasts}.

Ensemble simulation can be described as a simulation method that executes multiple simulations with different initial conditions, and each simulation is called ensemble member (hereinafter, simply referred as {\it member}). Therefore, an output from an ensemble simulation can be considered as a set of traditional time-varying, multivariate volumetric data, proportionately to the number of members. As a result, detailed analysis of an ensemble simulation data may require a great deal of effort and time to understand the underlying data and to obtain scientific knowledge. Domain experts in meteorology have historically used basic statistical functions, such as the mean and variance, for analyzing ensemble simulation outputs. However, in such aggregated data analysis, it becomes difficult to evaluate the simulation behavior among the members as well as in comparison to the ground truth data (treated hereinafter as {\it true state} to follow the term used in meteorology) such as the observational data from PAWR radar. Even in a case where the mean of all members for a certain variable being close to the true state, there is no guarantee that all members have close values to the true state. To enable such analysis, this requires an analysis from different facets of the ensemble data such as {\it members}, {\it variables}, {\it space}, and {\it time}.

We focused on the Parallel Coordinates Plot (PCP) which is a popular visualization technique used for analyzing high-dimensional datasets to identify outliers and trends as well as to understand the interrelations among variables. Although PCP cannot handle the aforementioned four facets simultaneously, it is possible to cover two of them, for instance, by using parallel axes for representing variables and using polylines for representing members. To overcome this problem, coordinated linked views have usually been applied to enable visual analysis of other facets such as the spatio-temporal facet. It is also worth noting that PCP can suffer from visual cluttering due to the overplotting with the increase in the number of members represented as polylines. In this work, to overcome this problem, we propose the Angular-based Edge Bundled Parallel Coordinates Plot (APCP) which simplifies the visualization by using representative lines connected through scatter plots with angular distribution information of the line segments. The APCP can show an overview of the entire ensemble dataset while maintaining the correlation information between adjacent variables. We developed a visual analytics system with linked coordinated views (\autoref {fig:teaser}) for the visual analysis of large ensemble simulation data. We used a meteorological ensemble simulation, with 100 members, of a sudden torrential rainfall occurred at the Kobe city, which was carried out on the supercomputer Fugaku. In this simulation, PAWR  radar data with 30 second resolution was used for the data assimilation.

\section{Related Work}



Various techniques and methods for the visualization and visual analysis of ensemble simulation data have already been proposed so far as we can verify in an extensive survey carried by Wang~et~al.~\cite{ wang2018visualization}. It is worth noting that visualization of ensemble simulation data is also an important topic in the meteorological data analysis as we can verify in a more specific survey carried out by Rautenhaus~et~al.~\cite{rautenhaus2018survey}. Considering the multifaceted aspects of the ensemble simulation data, most of the visual analytics tools have usually coordinated multiple linked views to enable intuitive user interaction and interactive visual exploration. As an example, we can cite the Ovis, an integrated visualization system for the ensemble heightfield data with multiple linked views combining 2D and 3D views~\cite{hollt2014ovis}. Another example is the EnsembleGraph for analyzing ensemble simulation data focusing on the behavior similarities between members over space and time~\cite{shu2016ensemblegraph}. In this system, a graph-based visual representation is applied to visualize spatio-temporal regions with similar behaviors. There is also visual analytics system which uses glyph-based visual representation for comparing the members against ground truth data~ \cite{bock2015visual}. Another glyph-based visual representation was proposed in~\cite{jarema2015comparative} for the visual analysis of vector field ensembles to visualize the directional probabilistic density functions. A glyph-based approach based on labeling was also used on a visual analytics system for the comparative analysis of 2D functions ensembles~\cite{piringer2012comparative}. There is an isocontour-based visualization for analyzing the temporal growth of the uncertainty in ensembles of weather forecasts~\cite{ferstl2016time}.

Parallel Coordinates Plot (PCP) ~\cite{inselberg2005pcp} is a popular visualization technique used for analyzing multivariate high-dimensional data to identify outliers and trends as well as to understand the interrelations among the variables. An extensive survey covering a variety of PCP-based research can be found in~\cite{heinrich2013pcp}, and a user-centered evaluation of a variety of PCP-based techniques can be found in~\cite{johansson2016pcp}. In this section, we focused on PCP-based methods for the quantitative evaluations of the correlations between variables and members. There are some approaches focusing on minimizing the visual cluttering due to the overplotting caused by the increase on the number of polylines. Wang~et~al.~\cite{wang2016multi} minimized the visual cluttering while stacking multiple PCP by inserting additional smaller axes between the main PCP axes. As a result, variable information from a given member at different resolutions becomes possible to be analyzed at the same time. However, when performing comparative analysis among members, the number of members that can be compared is limited by the need to insert a lot of smaller axes between the axes of the PCP or to juxtapose the PCP. 

Kumpf~et~al.~\cite{kumpf2021visual} performed cluster analysis with an overview of variable information in PCP. Then, by juxtaposing violin plots for each member, they could overview the data distribution of variables for multiple members in a cluster. However, the juxtaposition of violin plots requires more effort to analyze the similarity of each variable across members. Hazarika~et~al.\cite{hazarika2016visualizing} visualized the relationship of the spatial distribution of a single variable among members, with the axes of the PCP representing individual members. However, since the subject is a single variable, a large number of PCP must be juxtaposed in multivariate data. Yuan~et~al.~\cite{yuan2009scattering} proposed the Scattering Points in Parallel Coordinates (SPPC) where the problem of visual cluttering is minimized by inserting 2D scatter plots, obtained by applying multidimensional scaling to the target multidimensional data between the selected axes, and by drawing a curve passing through these scatter plots. As a result, it becomes possible to generate a PCP with edge bundling effect by agglomerating high similarity data (lines) in high-order dimension. This work was inspired in this SPPC approach.




\section{Methodology}
In this section, we describe the foundations of our proposed work. We will firstly explain the ensemble data, and will explain the proposed Angular-based Edge Bundled Parallel Coordinate Plot (APCP) including the Angular Distribution Plot (ADP) used for the scatter plots. Finally, we will explain the developed visual analytic system for analyzing large ensemble simulation data.

\subsection{Ensemble Data}
\begin{figure}[tb]
    \centering
    \includegraphics[height=0.6\textheight,width=0.9\hsize,keepaspectratio]{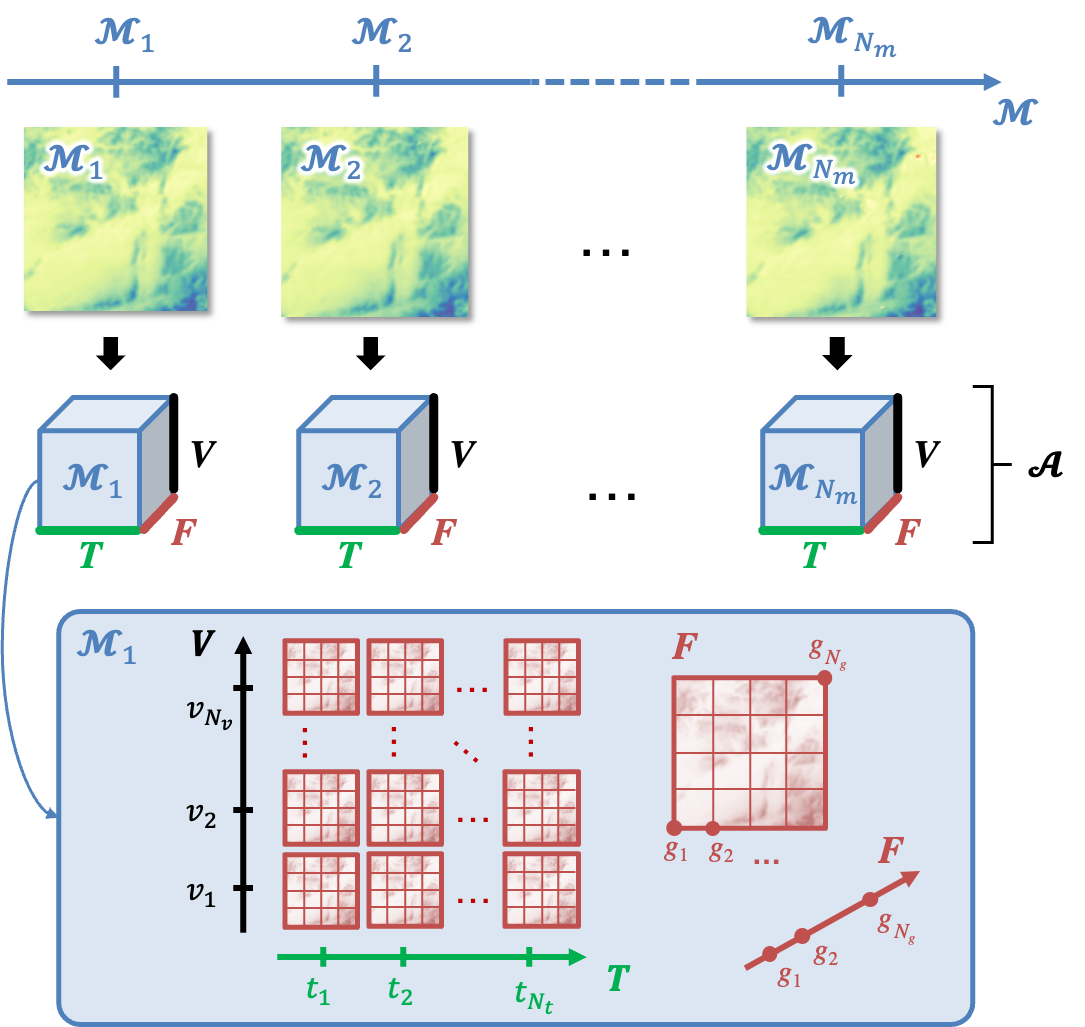}
    \caption{Ensemble data $\Tensor{A}$ with $N_m$ members. Each member $\Tensor{M}$ has a set of fields $\Mat{F}$ with time steps $N_t$. Numerical data $\Mat{V}$ composed of multiple variables is assigned to each grid point $g$.}
    \label{fig:ensembl_data}
\end{figure}

\begin{figure}[tb]
    \centering
    \includegraphics[height=0.6\textheight,width=1.0\hsize,keepaspectratio]{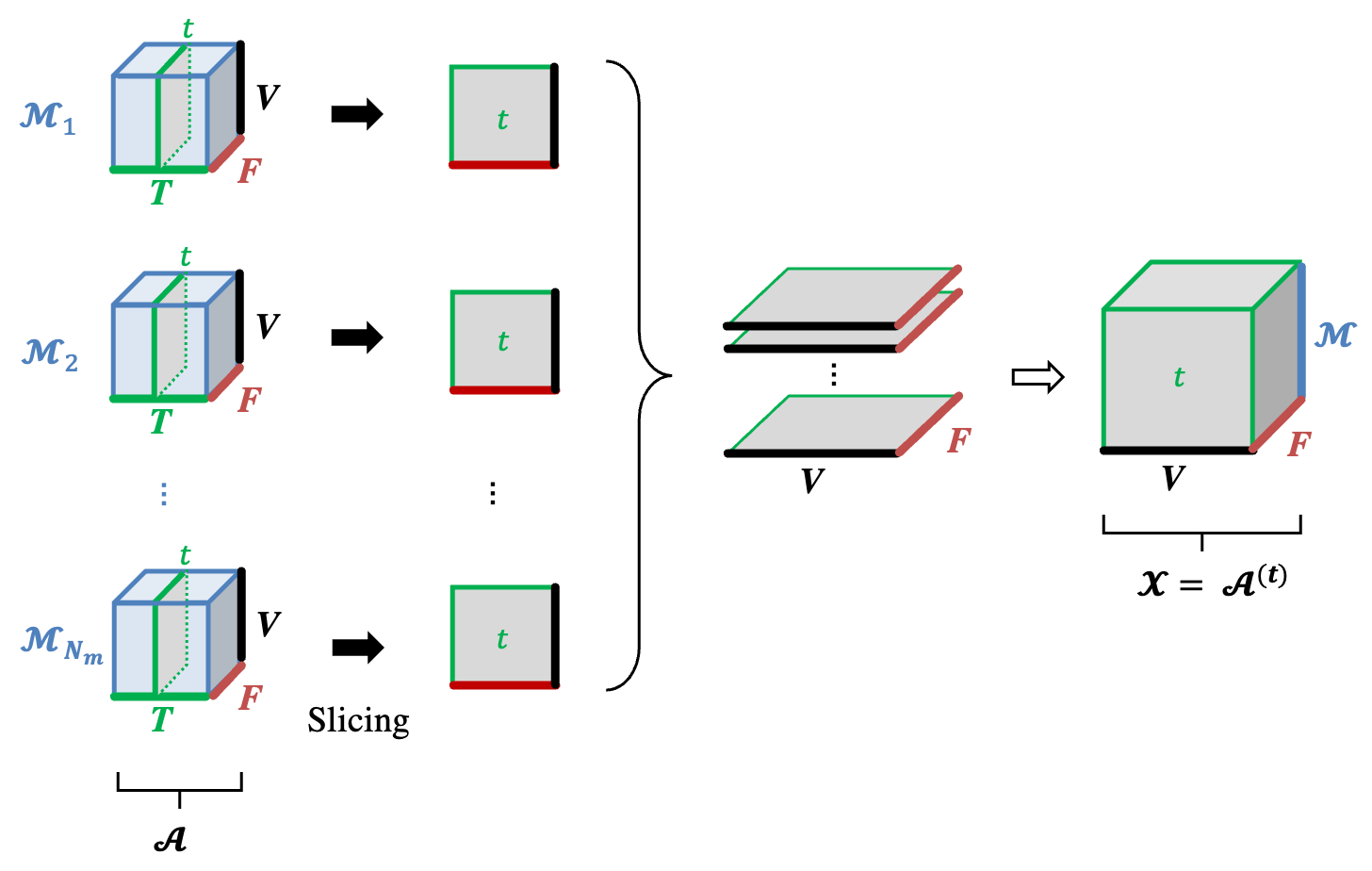}
    \caption{Slicing a 4th order tensor data $\Tensor{A}$ with respect to the $T$ axis, i.e. by fixing the time step $t$. The resulting sliced data $\Tensor{X} = \Tensor{A}^{(t)}$ will be a 3rd order tensor data.}
    \label{fig:slicing}
\end{figure}

Ensemble data treated in this work consists of a set of multiple multivariate time series (members) (\autoref{fig:ensembl_data}). This set $\Tensor{A}$ consisting of $N_ {m}$, number of members, can be expressed as $\Tensor{A} = \{\Tensor {M}_{1}, \Tensor {M}_{2}, \ldots, \Tensor{M}_{N_{m}} \}$, where $\Tensor{M}_i$ represents the $i$th member. The member $\Tensor{M}$ possesses the spatial information related to the underlying simulation (field), and can be expressed as a time series data. Assuming that the number of time steps in the simulation is $N_{t}$, and that the $i$th field is $\Mat{F}_{i}$, then the member $\Tensor{M}$ can be expressed as $\Tensor{M} = \{\Mat{F}_{1}, \Mat{F}_{2}, \ldots, \Mat{F}_{N_{t}}\}$. Considering that the time corresponding to the $i$th simulation time step is $t_{i}$, then the set of time steps $\Mat{T}$ can be expressed as $\Mat{T} = \{t_{1}, t_{2}, \ldots, t_{N_{t}}\}$. The field $\Mat{F}$ consists of a set of grid points defined in a Cartesian grid system. Assuming $N_ {g}$ as the number of grid points, then the field $ \Mat{F}$ can be expressed as $\Mat{F} = \{g_{1}, g_{2}, \ldots, g_{N_{g }} \}$, where $g_{i}$ represents the $i$th grid point. The set of variables, with the physical quantities calculated during the simulation steps, is assigned as numerical data at each of the grid points. Considering $N_{v} $ as the number of variables, the variable data $\Mat{V}_{g_{i}}$, on a grid point $g_{i}$, can be expressed as $\Mat{V} _{ g_{i}} = \{v_{1}, v_{2}, \ldots, v_{N_{v}} \}$, where $\Mat{V}_{j}$ represents the value of the $j$th physical quantity defined on the grid point $g_{i}$. As a result, the ensemble data based on member $\Tensor{M}$, field $\Mat{F}$, time $\Mat{T}$, and numerical data $\Mat{V}$ can be expressed as a fourth order tensor data $\Tensor{A}\ \in \mathbb{R}^{N_m \times N_g \times N_t \times N_v}$. The main objective of this representation is to facilitate the implementation and processing through a combination of simple data manipulations instead of a strict mathematical meaning. For the visual analysis, we assume that the time $t$ is defined in advance, that is, fixed prior to the analysis. Therefore, the target data for analysis (\autoref {fig:slicing}) is a cubic tensor data obtained by stacking the sliced data at time $t$ with respect to the time axis $T$ of each member, and can be represented as $\Tensor{A} ^ {(t)}\ \in \mathbb {R} ^ {N_m \times N_g \times N_v}$.

\subsection{Angular-based Parallel Coordinates Plot (APCP)}
This section describes our PCP-based approach to analyze the aforementioned cubic tensor data for a given time $t$ defined a priori by the user. We will firstly explain how PCP is used to visualize such multidimensional data, and then will explain the utilized approach for simplifying the visualization by using representative B\'{e}zier curves connecting the averaged line segments through the scatter plots representing the angular distribution information of the original line segments. This simplified visual representation with edge bundling effect is capable of showing an overview of the data while maintaining the correlation pattern information among adjacent parallel axes.

\subsubsection{PCP with Averaged Line Segments}

\begin{figure}[b]
    \centering
    \includegraphics[height=0.6\textheight,width=1.0\hsize,keepaspectratio]{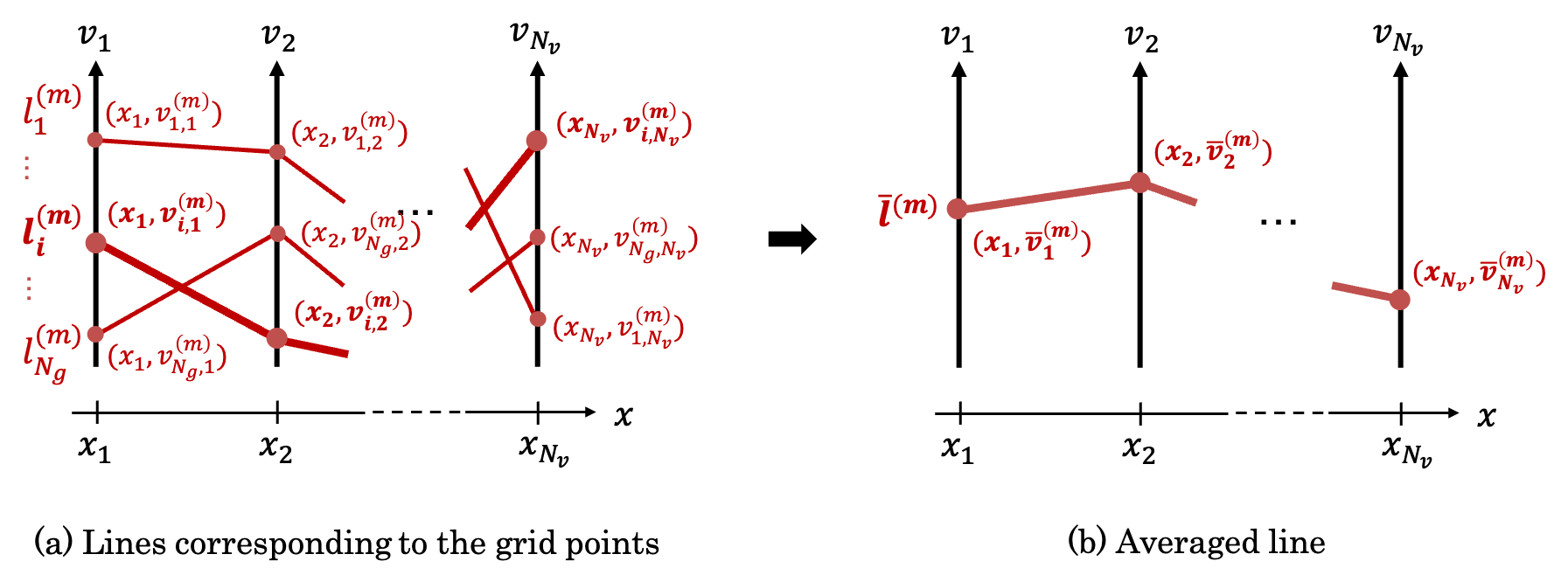}
    \caption{Plotting a representative line segment $\Mat{l}'^{(m)}$ for the $m$-th member passing through the mean values $\overline{v}_{i}^{(m)}$ for each of the variables $v_i$.}
    \label{fig:line_averaging}
\end{figure}

Since traditional PCP can simultaneously cover two dimensions of a given data, we used the parallel axes for representing the variables and the polylines for representing the members. For this purpose, as shown in \autoref{fig:pcp}, the cubic order tensor data $\Tensor{X} = \Tensor{A} ^ {(t)}$, obtained from the user defined time $t$, is sliced with respect to the member axis $\Tensor{M}$. Considering $\Tensor{X}_{m}$ as the sliced data corresponding to the $m$th member, this will correspond to a quadratic tensor data $\Tensor{X}_{m}\ \in \mathbb{R}^{N_g \times N_v}$, and will corresponds to an $N_g \times N_v$ matrix.  As shown in the right side of the figure, by using a polyline for each $\Tensor{X}_{m}$ to connect the member values for each of the variable axes $\{v_{1}, v_{2}, \ldots, v_{N_v} \}$ from $\Mat{V}$, we can obtain an overview of the relationships between the variables contained in $\Tensor{M}$. If we represent the horizontal axes of this parallel coordinates as $\{x_{1}, x_{2}, \ldots, x_{N_v} \}$, then the polyline $\Mat{l}_i^{(m)}$, corresponding to the grid points $g_i^{(m)}$, described in the $\Tensor{M}_{m}$, can be expressed as follows:

\begin{figure}[tb]
    \centering
    \includegraphics[height=0.6\textheight,width=1.0\hsize,keepaspectratio]{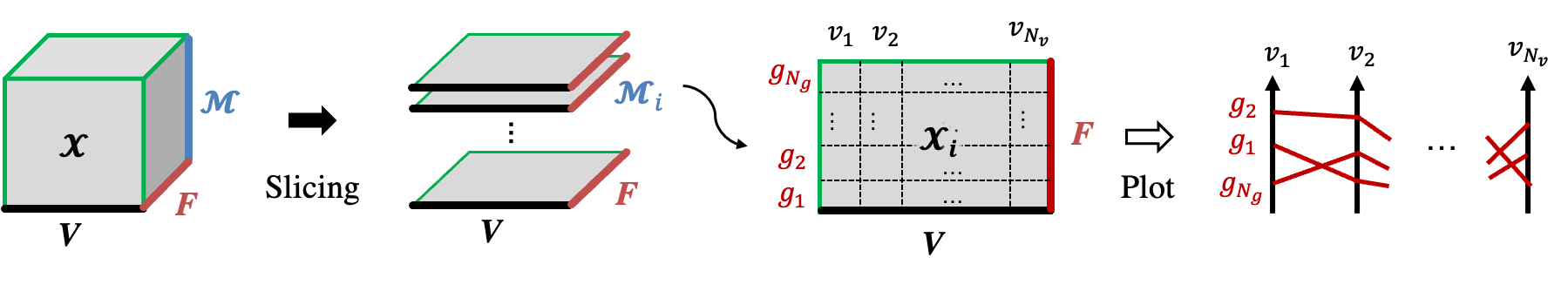}
    \caption{Parallel coordinates plot for the $\Tensor{X}_i$ sliced from the $\Tensor{X}$ at the position corresponding to the $i$-th member $\Tensor{M}_i$. Multiple line segments for each grid point $g$ will be plotted in the parallel coordinates composed of axes with respect to the variables $v$.}
    \label{fig:pcp}
\end{figure}


\begin{equation} \label{eq:lm}
    \Mat{l}_i^{(m)} = \{ (x_1, v_{i,1}^{(m)}), (x_2, v_{i,2}^{(m)}), \ldots, (x_{N_v}, v_{i,{N_v}}^{(m)}) \}
\end{equation}
Here, $v_{i, j}^{(m)}$ corresponds to the $j$th physical quantity value defined on the grid point $g_{i}^{(m)}$, and $(x_j, v_{i, j}^{(m)}) $ represents a point on the $j$th parallel axis. As shown in \autoref{fig:line_averaging} (a), the polyline $\Mat{l}_i^{(m)}$ will be drawn by connecting the line segments between these points defined on the parallel axes.

However, when drawing the $\Tensor{M} _{m}$ as polylines $\{\Mat{l}_i^{(m)} | 1 \leq i \leq N_g, 1 \leq m \leq N_m \}$ on the parallel coordinates, the rendering cost can be negatively affected with the increase in the number of grid points $N_g$, thus impacting the interactive visual exploration. In addition, visual cluttering due to the overplotting of line segments may compromise the visual analysis. To minimize such problem, in our proposed method, we used averaged line segments passing through the mean value for each of the variables, represented by the horizontal axes as shown in the \autoref{fig:line_averaging} (b).  Considering $\overline{v}_j^{(m)}$ as the mean value of the $j$th physical quantity value, contained in the $\Tensor{M}_m$, these representative polylines $\Mat{ l}'^{(m)}$ can be expressed as follows:



\begin{equation} \label{eq:l'm}
    \Mat{l}'^{(m)} = \{ (x_{1}, \overline{v}_{1}^{(m)}), (x_{2}, \overline{v}_{2}^{(m)}), \ldots, (x_{N_v}, \overline{v}_{N_v}^{(m)}) \}
\end{equation}
As a result, it becomes possible to minimize the visual cluttering problem, brought about by the increase in the number of polylines, by drawing solely the representative line segment $\{\Mat{l}'^{(m)} | 1 \leq m \leq N_m\} $ for each of the members. Although this enables to overview more efficiently the relationships among the members, the main drawback is the missing of correlation information among the variables such as those shown in \autoref{fig:corr_pattern}. In the next subsection, we will present the angular distribution plot used to overcome this problem.

\subsubsection{Angular Distribution Plot (ADP)}
\begin{figure}[t]
    \centering
    \includegraphics[height=0.6\textheight,width=1.0\hsize,keepaspectratio]{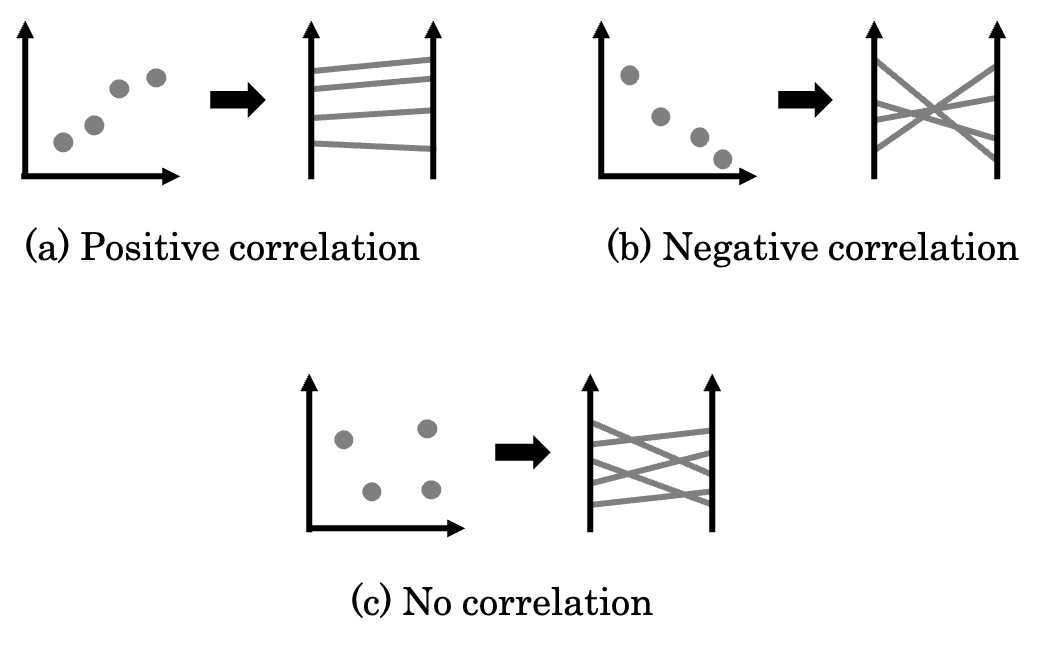}
    \caption{Correlation patterns inferred from the intersection patterns of line segments between adjacent parallel coordinate axes.}
    \label{fig:corr_pattern}
\end{figure}

It is commonly possible to visually infer the correlations between adjacent variables in the PCP by analyzing the intersection patterns of the line segments between the adjacent parallel axes. \autoref{fig:corr_pattern} shows the well-known point-line duality where a scatter plot (point) in Cartesian coordinates becomes a line in the PCP, and presents the possible correlation patterns. We can infer a positive correlation when there is no intersection between the axes as shown in \autoref{fig:corr_pattern} (a). On the other hand, we can infer a negative correlation when there are many line segment intersections on the center region of the adjacent axes as shown in \autoref{fig:corr_pattern} (b). Finally, we can infer that there is no correlation when the line segments intersect in a random manner without a certain pattern as shown in \autoref{fig:corr_pattern} (c). Trying to efficiently express these intersection patterns, we take into consideration the angles of the line segments between adjacent parallel axes.

\begin{figure}[b]
    \centering
    \includegraphics[height=0.6\textheight,width=0.6\hsize,keepaspectratio]{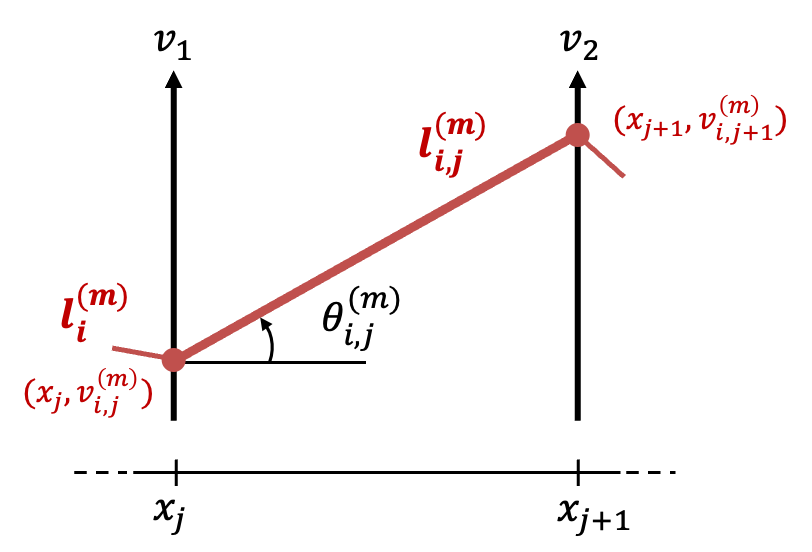}
    \caption{Angle $\theta _{i,j}^{(m)}$ of the line segment $\Mat{l}_{i,j}^{(m)}$ for the $i$-th grid point between the adjacent axes $x_{j}$ and $x_{j+1}$.}
    \label{fig:line_angle}
\end{figure}

As shown in \autoref{fig:line_angle}, to obtain the angle of a polyline $\Mat{l}_i^{(m)}$ (\autoref{eq:lm}), we focus on each of the line segments $\Mat{l}_{i, j}^{(m)}$ between the corresponding adjacent axes $x_j, x_{j + 1}$. Considering $(x_j, v_ {i, j} ^ {(m)})$ as one of the extremes of this partial line segment on the $x_j$ axis, and $(x_{j +1}, v_ {i, j + 1}^{(m)})$ as the other extreme on the $x_{j + 1}$ axis, the angle $\theta_{i, j}^{(m)} $ for this particular line segment $\Mat{l}_{i, j}^{(m)}$ can be obtained as follows:

\begin{equation}\label{eq:theta}
    {\theta _{i,j}}^{(m)} = \arctan ( {v'_{i,j+1}}^{(m)} - {v'_{i,j}}^{(m)})
\end{equation}
Here, ${v'_{i, j}}^{(m)}$ represent the value obtained by normalizing ${v_{i, j}}^{(m)}$ by using the maximum and minimum values of the entire member values, and by assuming that the distance between adjacent axes is 1, that is, $| x_j - x_{j + 1} | = 1$. From the obtained angle values, by using the Eq.~\ref{eq:theta}, the mean $\bar {{\theta}_j}^{(m)}$ and variance $\hat{{\theta}_j}^{(m)}$ corresponding to  each of the members can be obtained as follows:

\begin{eqnarray}
\label{eq:theta_mean}
    \bar{\theta}_j^{(m)} & = & \frac{1}{N_g} \sum _{i=1}^{N_g} \theta _{i,j}^{(m)}\\
\label{eq:theta_var}
    \hat{\theta}_j^{(m)} & = & \frac{1}{N_g} \sum _{i=1}^{N_g} (\theta _{i,j}^{(m)} - \bar{\theta}_j^{(m)})^2
\end{eqnarray}

\begin{figure}[b!]
    \centering
    \includegraphics[height=0.6\textheight,width=1.0\hsize,keepaspectratio]{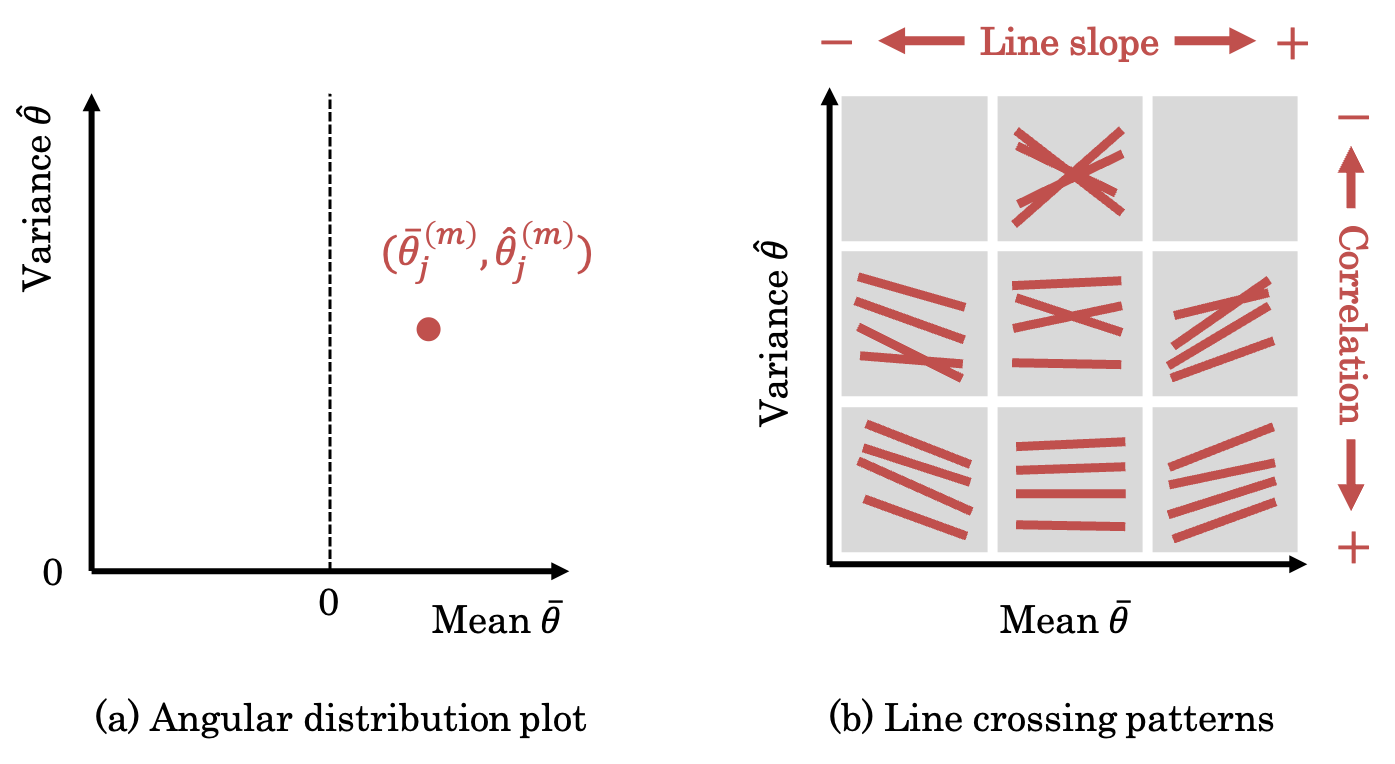}
    \caption{Angular distribution plot and the intersection patterns of the line segments. (a) shows the angular distribution plot generated on the 2D grid composed of the mean axis $\bar {\theta}$ and the variance axis $\hat {\theta}$. (b) shows the relationships between the intersection patterns and the corresponding plotting region. The horizontal axis corresponding to the mean indicates the line slope, and the vertical axis corresponding to the variance indicates the correlation between the variables.}
    \label{fig:angle_dist}
\end{figure}

As shown in \autoref{fig:angle_dist} (a), these mean and variance values, obtained respectively from \autoref{eq:theta_mean} and \autoref{eq:theta_var}, are used as the tuples $(\bar {\theta}_j^{(m)}, \hat {\theta}_j^{(m)})$ for generating the Angular Distribution Plot (ADP). In the ADP, the horizontal axis represents the mean, and the vertical axis represents the variance. We assume that the axes of the 2D grid are set in such a way where the mean value 0 ($\bar{\theta} = 0$) is placed at the center of the horizontal axis, and the variance value 0 ($\hat{\theta} = 0$) is situated at the bottom of the vertical axis. As a result, an ADP reflecting the intersection patterns between adjacent axes will be generated by $N_m$ scatter plots at each of the $N_v -1$ regions between adjacent axes.

\autoref{fig:angle_dist} (b) shows the relationships between the intersection patterns and the corresponding plotting regions. The horizontal axis (mean) indicates the slope or tendency of the line segments, that is, small mean value indicates that many line segments have negative slope, and on the other hand, large mean value indicates that many line segments have positive slope. The vertical axis (variance) indicates the correlation between the variables. That is, the smaller the variance, the less the intersection between line segments thus more positive will be the correlation. On the other hand, the larger the variance, the more the intersections between line segments near the central region, thus more negative will be the correlation. 

\subsubsection{Edge Bundling using Angular Distribution Plots}
\begin{figure}[b!]
    \centering
    \includegraphics[height=0.6\textheight,width=1.0\hsize,keepaspectratio]{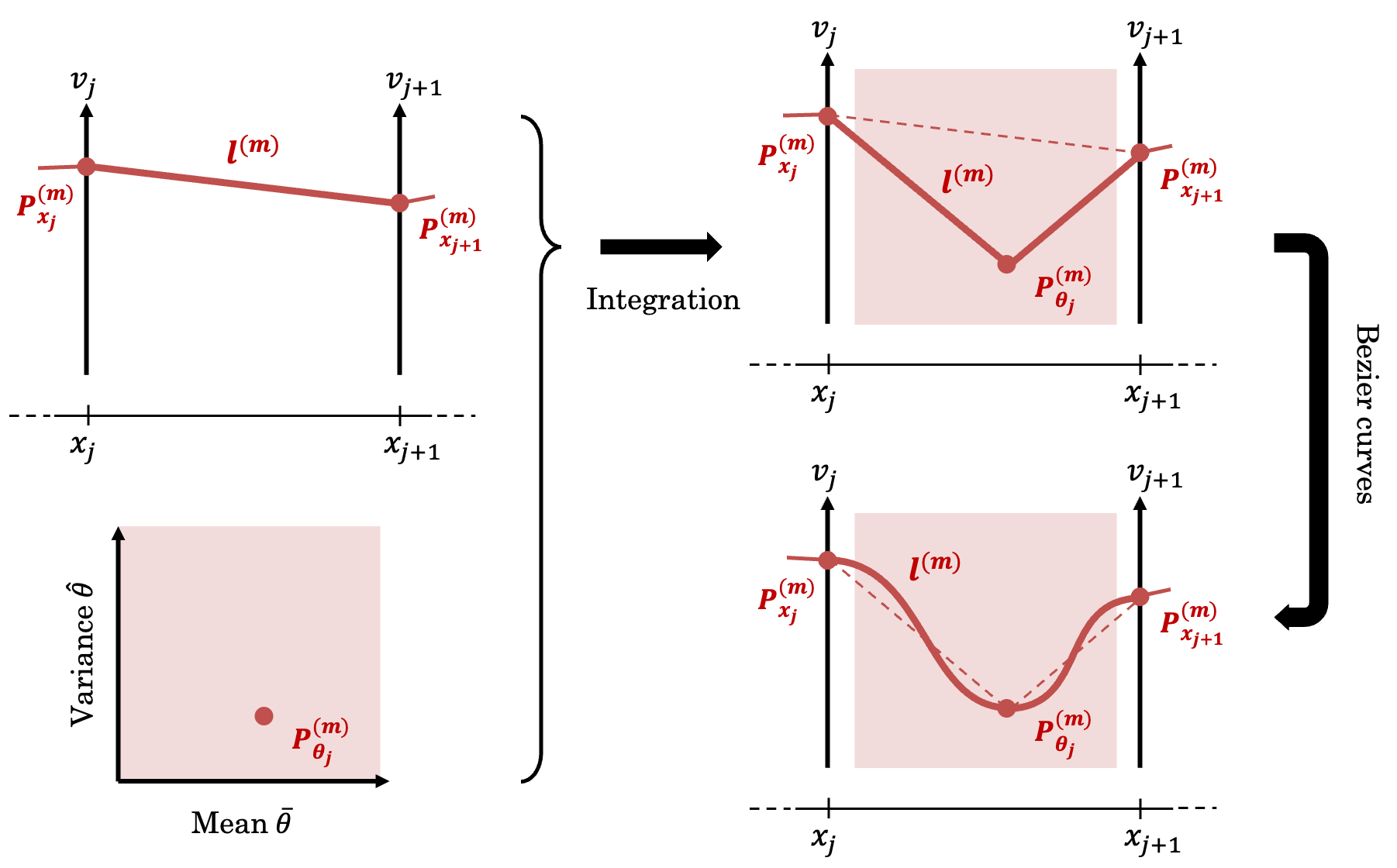}
    \caption{Drawing B\'{e}zier curves with $P_{x_j}^{(m)}$, $P_{ x_{j + 1}}^{(m)}$ and $P_{\theta_j}^{(m)}$. The points $P_{x_j}^{(m)}$ and $P_{ x_{j + 1}}^{(m)}$ are the starting and ending points of the line segment $\Mat{l}^{(m)}$, respectively, and the $P_{\theta_j}^{(m)}$ is the intermediate point as being the angular distribution plot between the adjacent axes $x_j$ and $x_{j + 1}$.}
    \label{fig:bezier_curves}
\end{figure}

The averaged line segments and the scatter plot (ADP) corresponding to each of the members will be converted to a B\'{e}zier curve connecting the extremes of the line segment passing through the scatter plot as shown in \autoref{fig:bezier_curves}. The step-by-step procedure for obtaining this B\'{e}zier curve is as follows:

\begin{enumerate}[leftmargin=1.1cm, label=Step \arabic*:]
    \item Insert the scatter plots (ADP) at  each of the corresponding regions between the parallel axes.
    \item Consider $P_{x_j}^{(m)}$ and $P_{ x_{j + 1}}^{(m)}$ as the endpoints of the  line segment, corresponding to a member $m$, between the adjacent axes $x_j$ and $x_{j + 1}$.
    \item Consider $P_{\theta_j}^{(m)}$ as the scatter plot, corresponding to a member $m$, between the adjacent axes $x_j$ and $x_{j + 1}$.
    \item Draw a B\'{e}zier curve connecting these three points $P_{x_j}^{(m)}, P_{\theta_j}^{(m)}, P_{x_{j + 1}}^{(m)}$ with the point $P_{\theta_j}^{(m)}$ serving as the intermediate point. At this time, the segments $P_{x_j}^{(m)}, P_{\theta_j}^{(m)}$ and $P_{\theta_j}^{(m)}, P_{x_{j + 1}} ^ {(m)}$ will be connected by a cubic B\'{e}zier curve, and the control points should be appropriately set in a way that the neighboring region of $P_{\theta_j}^{(m)}$ is continuously differentiable.
    \item Execute the above four steps for all members and interaxis regions.
\end{enumerate}

\begin{figure}[tb]
    \centering
    \includegraphics[height=0.6\textheight,width=1.0\hsize,keepaspectratio]{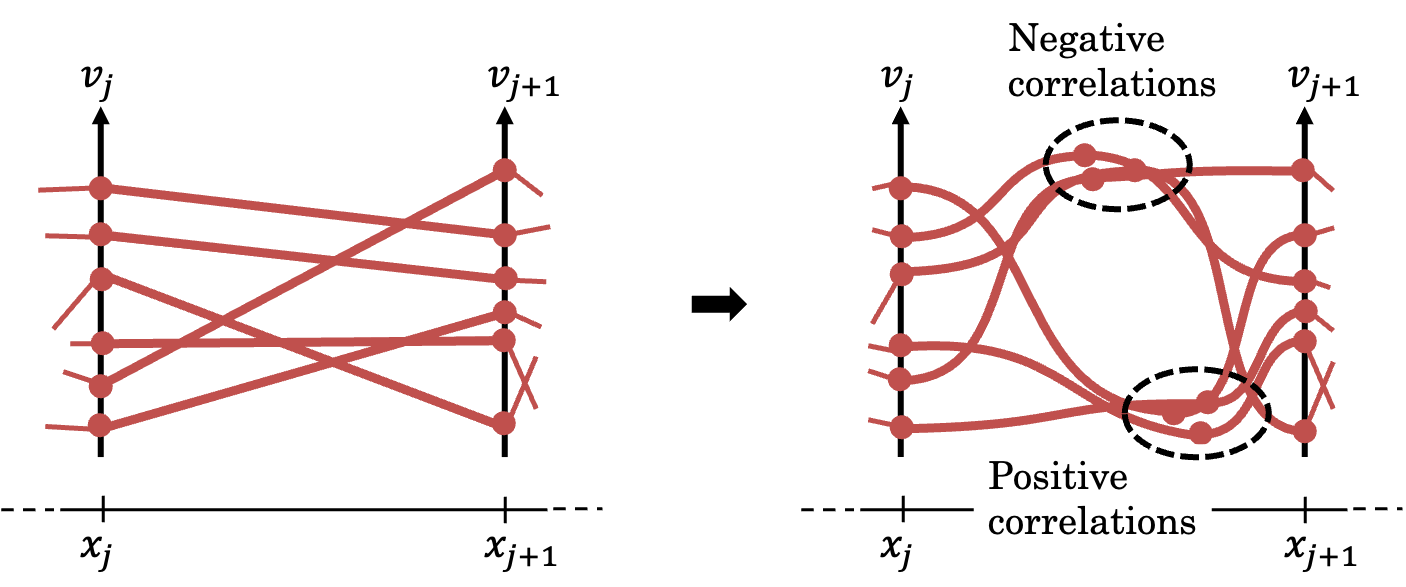}
    \caption{Bundling effect based on the similarity of members indicated by the ADP.}
    \label{fig:bundling}
\end{figure}

The obtained APCP will be composed of representative line segments, for each of the members, drawn as B\'{e}zier curves passing through the scatter points with the angular distribution information. As mentioned before, our approach was inspired in the Scattering Points in Parallel Coordinates (SPPC) proposed by Yuan~et~al.~\cite{yuan2009scattering}. In our proposed method, instead of using dimensionality-reduced scatter plot, we used the ADP to be inserted in the interaxis regions to draw the representative curves. As shown in \autoref{fig:bundling}, this visual representation produces the edge bundling effect based on the similarity among the members. This simplified view is capable of showing an overview of the ensemble data while maintaining the correlation pattern information among the adjacent parallel axes.

\subsection{Visual Analytics System}
We implemented a visual analytics prototype system for evaluating the proposed APCP approach. It is worth noting that as other visual analytics system for ensemble simulation data, some additional views, in the form of coordinated linked views, are used to assist the interactive visual analysis of a given ensemble simulation data for a pre-defined simulation time step $t$. \autoref{fig:teaser} shows an overview of the visual analytics system composed of the main APCP view in addition to the ADP view, a  Binning Parallel Coordinate Plot (BPCP) view, and a Cross Sectional Plot (CSP) view. 

In the APCP view (\autoref{fig:teaser}(a)), an overview of the ensemble data including all members and their variables will be presented. On the APCP view, users are able to interactively select a member (represented as B\'{e}zier curves) of interest, and this selected member will be displayed in red color, in contrast to other members displayed in the default blue color. In addition, by selecting an interaxis region of interest, the corresponding scatter plot  will be displayed in the ADP view with the member of interest plotted in red color as shown in \autoref{fig:teaser}(b).



In the BPCP view (\autoref{fig:teaser}(c)), all variables for the selected member of interest will be displayed. To minimize the visual cluttering problem when the amount of $N_g$ line segments becomes large, we adopted the multidimensional binning technique~\cite{heinrich2013pcp}. Although we used a pre-defined number of bins for each of the variables, that is, user-defined number of bins, it is worth noting that there exist approaches for estimating the optimal number of bins such as those based on the Sturges's formula~\cite{sturges1926choice}, Doane's formula~\cite{doane1985propagation}, Scott's formula~\cite{scott1979optimal}, or Freedman and Diaconis's formula~\cite{freedman1981histogram}~\cite{scott2015multivariate}. In the BPCP view, the bins with larger data will be displayed on the front, and users are able to interactively brush the range of the value distribution for an axis (variable) of interest.

The CSP view (\autoref{fig:teaser}(d)) shows the spatial distribution (field) of the user selected variable. This view shows the horizontal section of the volume data corresponding to the member selected in the APCP view. Users are able to interactively navigate through the cross sectional views by manipulating the slider presented in the upper right position.


\begin{figure*}[htb!]
    \centering
    \includegraphics[height=0.6\textheight,width=1.0\hsize,keepaspectratio]{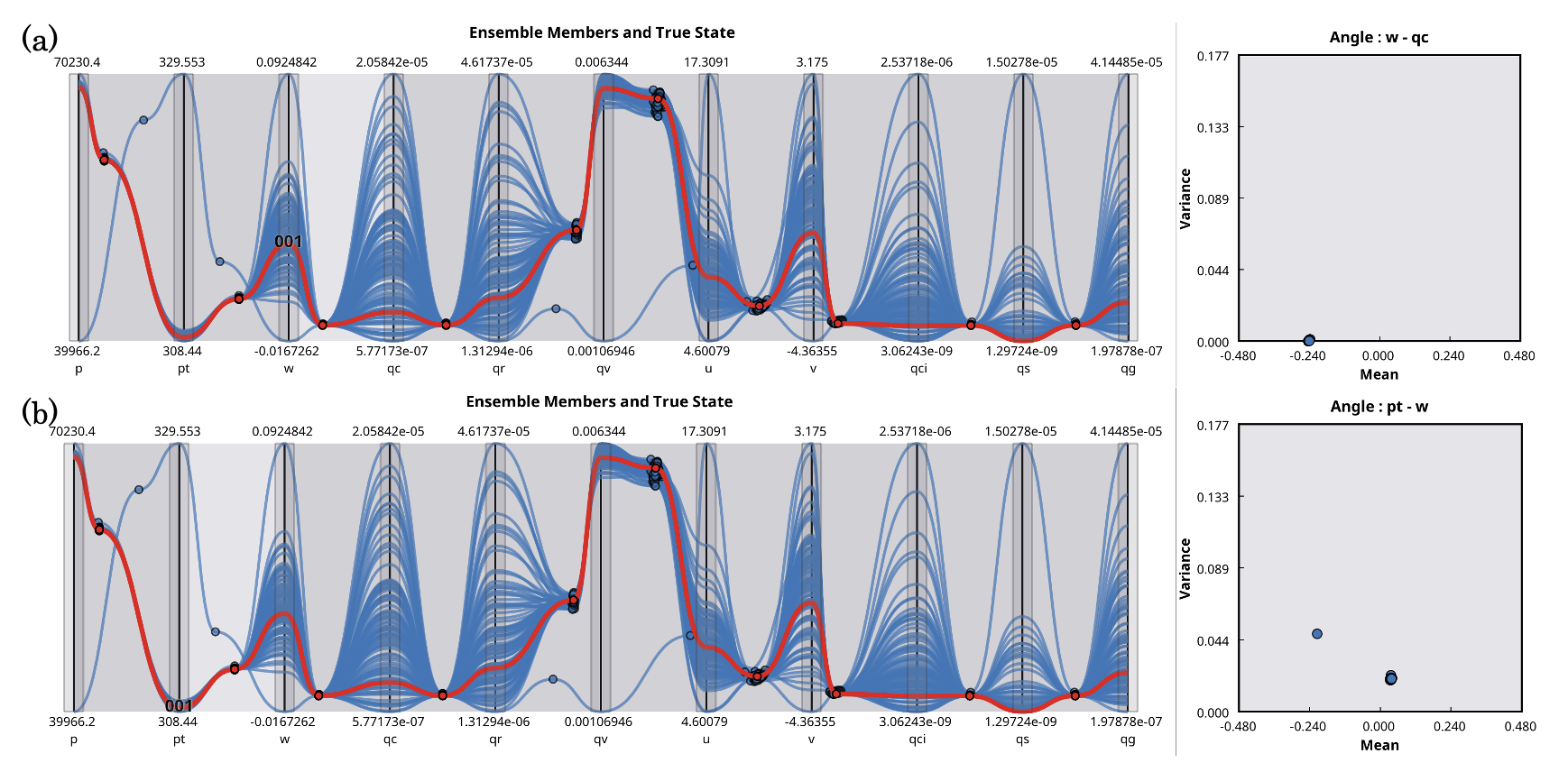}
    \caption{Some APCP and ADP results for the meteorological ensemble simulations with 100 members and the true state: (a) shows a case focused on the interaxis region between $w$ and $qc$ axes, and (b) shows a case focused on the interaxis region between $pt$ and $w$ axes.}
    \label{fig:APCP_ADP}
\end{figure*}

\section{Experimental Evaluations and Results}
We implemented the visual analytics system by using C++ language, and used the Kyoto Visualization System (KVS)~\cite{sakamoto2015kvs}, a visualization application development platform, to implement the necessary underlying functionalities. We also used the Qt6~\cite{qt6} for implementing the user interface (GUI) shown in \autoref{fig:teaser}. For the evaluations, we used a single node from  the 8-node x86-based Server, with each node consisting of Dual Intel Xeon Gold 6238R CPU (28 cores / 2.20 GHz), 384 GB of DRAM, and an NVIDIA Quadro RTX 8000 GPU.

\subsection{Data}


In the experiment, we used a meteorological ensemble simulation, with 100 members, of a torrential rainfall occurred at the Kobe city on September 11, 2014 ~\cite{maejima2017}. The simulation data has grid points of $160 \times 160$ with $500m$ horizontal resolution, and $48$ layers with $350m$ mean vertical resolution. In this simulation, data assimilation with 30 second update cycle was performed by using phased array weather radar data from 08:00 JST (Japan Standard Time) to 08:30 JST. Following the initial data assimilation stage, a 15 minute numerical weather forecast, starting from 08:15 JST, was carried out and generated simulation output at every 30 seconds. The grid point data corresponds to the ensemble mean of the data assimilated values, and the Ensemble Kalman Filter was employed as the data assimilation scheme, which assumes a Gaussian probability distribution in the estimation of each ensemble state. Analysis data with optimal values obtained from the data assimilated result was assumed to provide the best estimation of the atmospheric state, and was utilized in the experiment as the true state. The simulation output has a total of 11 variables: east-west-component (x-component) of the horizontal wind velocity (u); north-south-component (y-component) of the horizontal wind velocity (v); vertical component of the wind velocity (w); potential temperature (pt); pressure (p); mixing ratio of water vapor (qv), mixing ratio of cloud water (qc), mixing ratio of rainwater (qr), mixing ratio of cloud ice (qci), mixing ratio of snow (qs), and mixing ratio of graupel (qg). The mixing ratio is defined as the mass of hydrometeor divided by 1 kg dry air. These variables were represented as floating point values (4 bytes), and as a result, each member's volume data has around 54 MB ($160 \times 160 \times 48 \times 11 \times 4$) with a total of 5.4 GB (100 members and 1 true state) per time step (total of 31).    

\begin{figure}[b!]
    \centering
    \includegraphics[height=0.6\textheight,width=1.0\hsize,keepaspectratio]{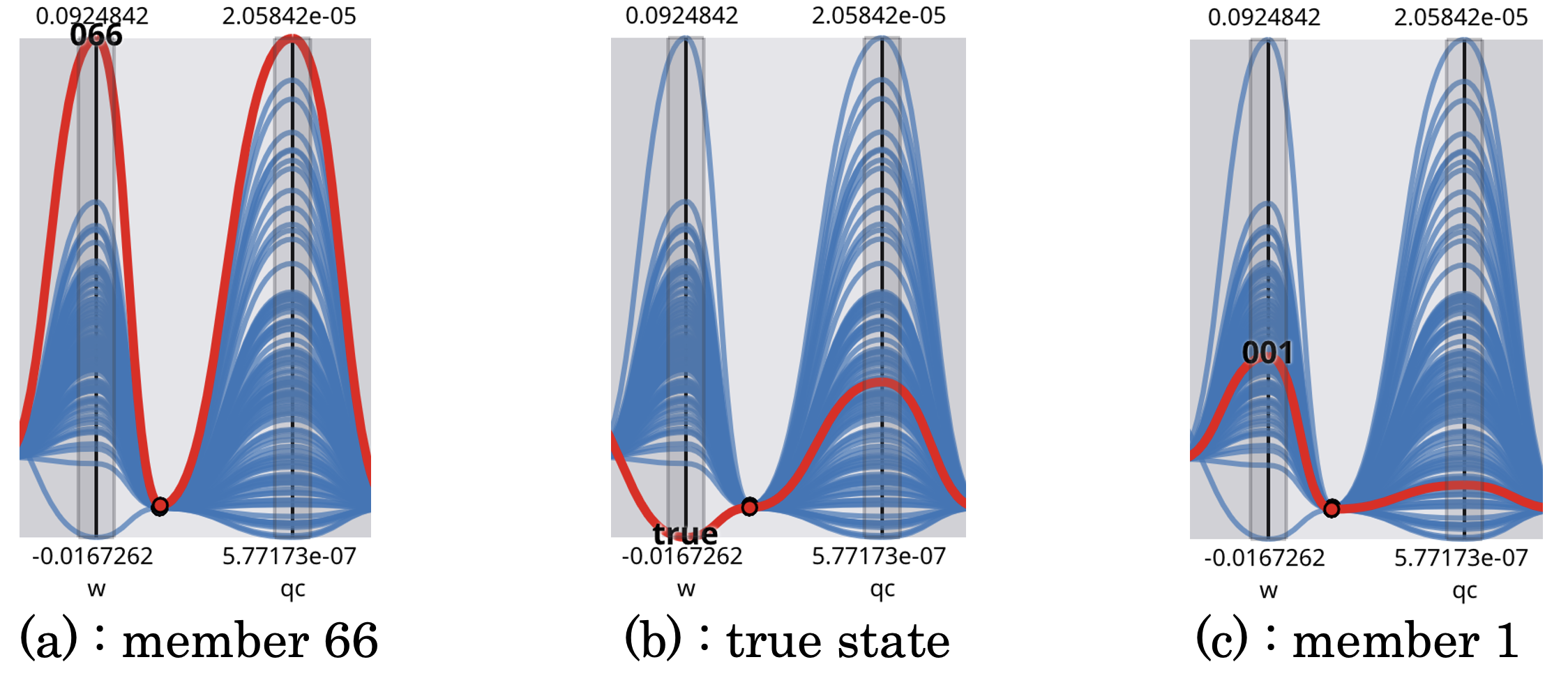}
    \caption{APCP views focusing on the $w-qc$ region. Red colored curves represent the member 66 (a), true state (b), and member 1 (c).}
    \label{fig:wqc_APCP}
\end{figure}

\subsection{Case Studies}

We conducted two case studies with the domain expert who carried out the meteorological ensemble simulation, and assisted the development and evaluation of the proposed approach since the beginning. Previous analysis using the ensemble mean (of the 100 members) was already carried out by the domain expert, and presented in~\cite{maejima2017}. However, since the detailed analysis using the entire set of members was not carried out, two cases were selected for the evaluations. One of them, shown in \autoref{fig:APCP_ADP} (a), focuses on the interaxis region between $w$ and $qc$ axes, representing respectively the vertical component of the wind velocity and the mixing ratio of cloud water. The $qc$ is the condensation of water vapor ($qv$) in the atmosphere, and it is strongly connected with the precipitation procedure. Moreover, the intense upward motion ($w$) contributes to accelerate the translation from $qv$ to $qc$. Hence, this relationship indicates the potential for the occurrence of the heavy rainfall. Focusing on \autoref{fig:APCP_ADP} (a), we can observe the presence of members plotted far away from each other on the $w$ axis. Here, we can also observe a small dispersion on the angular distribution since the scatter plots are concentrated in a single region on the ADP view. Another case study, shown in \autoref{fig:APCP_ADP} (b), focuses on the interaxis region between the $pt$ and $w$ axes, representing respectively the potential temperature and the vertical component of the wind velocity. The relationship between $pt$ and $w$ is  associated with the severe rainfall. Generally, in a developing period of the convection, intense $w$ is often found and is maintained by an unstable stratification characterized by a vertical variation of $pt$. Thus, this relationship is noteworthy to determine the activation of the convection, which leads to the severe rainfall. Here, we can observe a member on the $pt$ axis placed far away from others on the APCP view (\autoref{fig:APCP_ADP} (b)). We can also observe that the corresponding scatter plot is also placed separately from other plots on the ADP view. For these two cases, we will discuss in the next section the detailed analysis results obtained from the visualization system. 

\subsubsection{Case 1: Vertical Component of the Wind Velocity $w$ and Mixing Ratio of Cloud Water $qc$}
\label{case_1}


Focusing on the APCP and ADP results shown in~\autoref{fig:APCP_ADP} (a), and looking to the $w$ axis on the APCP, we can verify that the largest and smallest valued members are plotted distant from the rest of the members. As shown in \autoref{fig:wqc_APCP} (a) and (b), we can observe that the member with the largest mean value on the $w$-axis is the member 66, and the member with the smallest mean value is the member corresponding to the true state. Now, when looking to the ADP view in \autoref{fig:APCP_ADP} (a), we can verify that all scatter plots are concentrated on a single region. In other words, we can infer that the intersection pattern for all members in the interaxis region of $w$ and $qc$ axes are similar. Considering that the concentrated scatter plots are placed in a region where the mean is negative and the variance is near 0, we can infer a positive correlation between $w$ and $qc$, and the existence of several line segments with negative slope.



\begin{figure}[bt!]
    \centering
    \includegraphics[height=0.6\textheight,width=1.0\hsize,keepaspectratio]{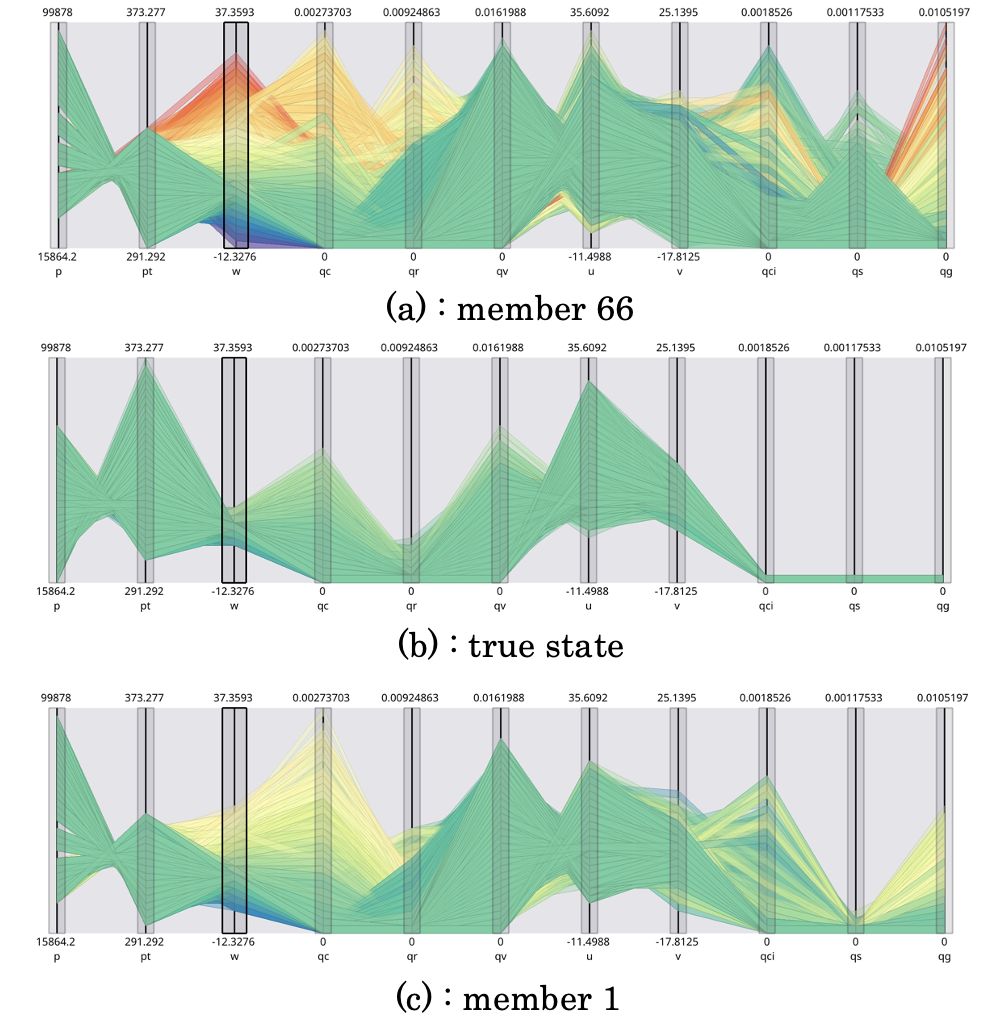}
    \caption{BPCP views focusing on the $w$-axis. Member 66, True state member, and  member 1 are selected in (a), (b), and (c).}
    \label{fig:wqc_BPCP}
\end{figure}

\begin{figure}[h!]
    \centering
    \includegraphics[height=0.6\textheight,width=1.0\hsize,keepaspectratio]{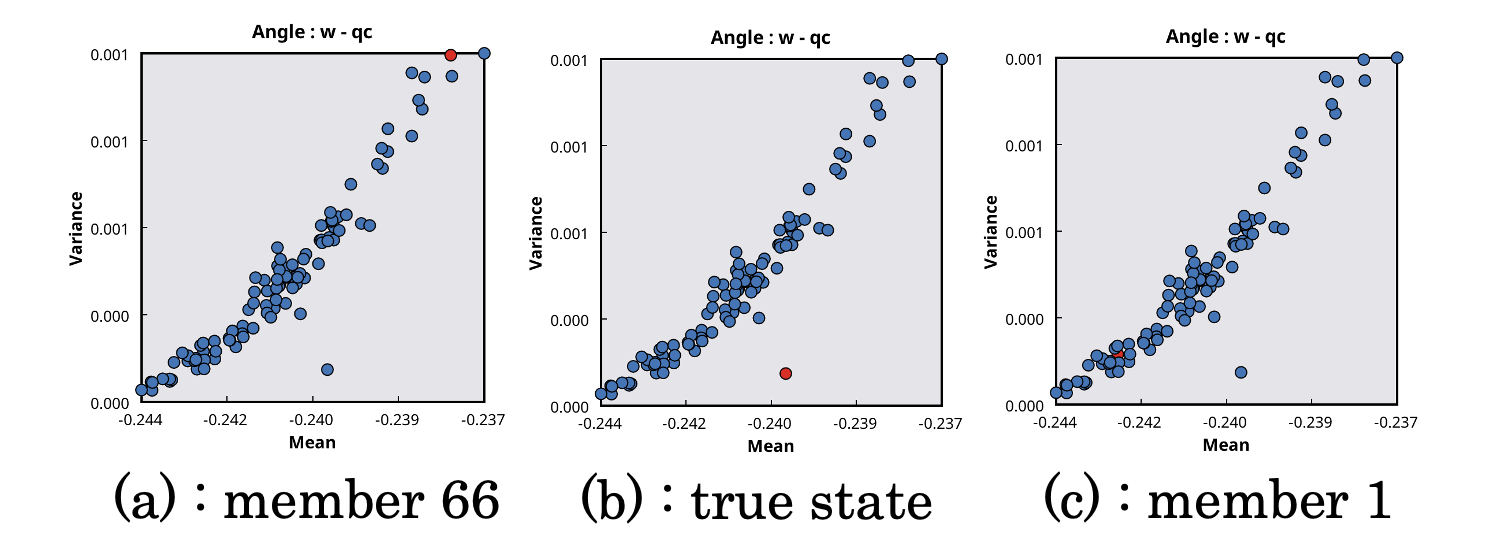}
    \caption{ADP results with rescaled axes range. The member 66, the true state, and the member 1 are selected in (a), (b), and (c).}
    \label{fig:wqc_ADP}
\end{figure}

\begin{figure}[h!]
    \centering
    \includegraphics[height=0.3\textheight,width=1.0\hsize,keepaspectratio]{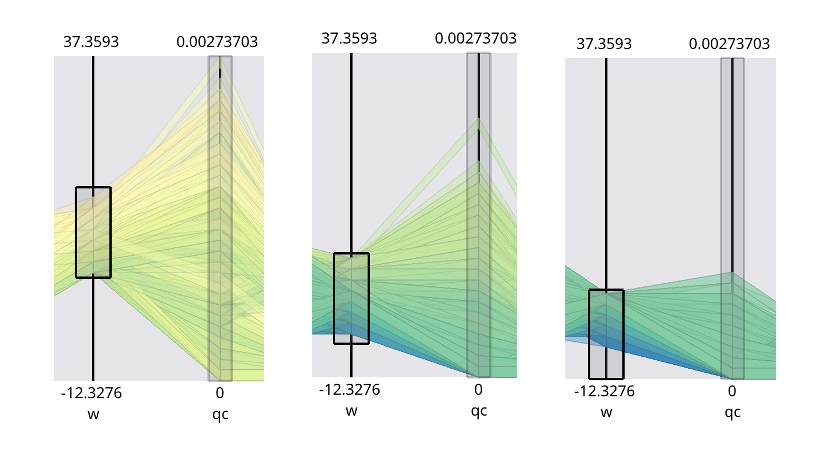}
    \caption{Changing the drawing range, via brushing, of the $w$-axis in the BPCP for the member 1.}
    \label{fig:wqc_ranged_BPCP}
\end{figure}

\begin{figure}[h!]
    \centering
    \includegraphics[height=0.6\textheight,width=1.0\hsize,keepaspectratio]{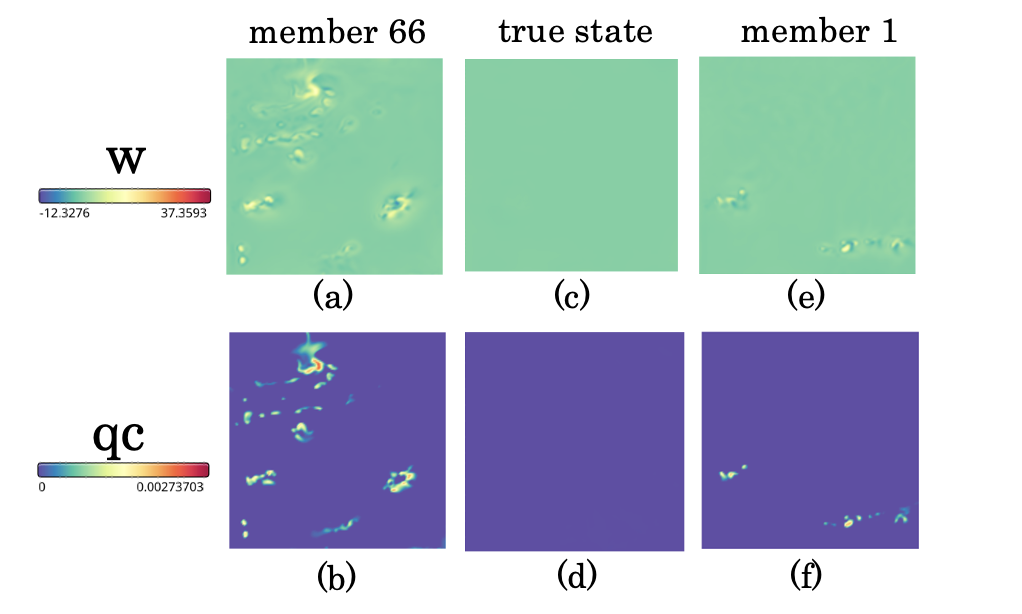}
    \caption{Cross-sectional views for the variables $w$ and $qc$.}
    \label{fig:wqc_slice}
\end{figure}

Regarding the positive correlation in the interaxis region between the variables $w$ and $qc$ among all members, the domain expert explained that this can occur when the convection is in the developing stage. Since the positive correlation trend is observed to all the members, then it can be considered that they are all contributing to activate the convection. The domain expert also explained that the fact of the true state having smaller value than all mean values of variable $w$, was previously noted, and  discussed in~\cite{maejima2017}. To further investigate this case, we focused on three members as shown in \autoref{fig:wqc_APCP}. In addition to the members 66 and of the true state, we used the member 1 which has a mean value close to most of other members. \autoref{fig:wqc_BPCP} shows the BPCP results for the selected three members, and we can verify that intersection patterns are dissimilar. Focusing on the $w$ axis, we can verify the existence of larger values in the member 66 compared to other two members (\autoref{fig:wqc_BPCP} (a)). We can verify that most of the values on the $w$ axis of the true state member are in the small value range (\autoref{fig:wqc_BPCP} (b)). We can verify that member 1 has values on the $w$ axis in the range between the members 66 and of the true state (\autoref{fig:wqc_BPCP} (c)). These observations are consistent with the APCP results where the mean value of $w$ for the member 66 is larger than of other members, and the mean value of $w$ for the member representing the true state is smaller than of other members.


Focusing on the interaxis region between the $w$ and $qc$ axes of \autoref{fig:wqc_BPCP}, 
and considering that the scatter plots were concentrated into a single region of the ADP view (\autoref{fig:APCP_ADP} (a)), we can expect larger angular variance for the member 66, in comparison to other members. To verify this hypothesis, we rescaled the ADP axes, and we could confirm that the angular variance of member 66 is larger than the other two members (\autoref{fig:wqc_ADP}). To verify the inferred correlation, we changed via brushing the drawing range of the $w$-axis in the BPCP for the member 1 (\autoref{fig:wqc_ranged_BPCP}). As a result, we could observe that most of the line segments intersect outside the parallel axes, thus it is possible to infer that $w$ and $qc$ have a positive correlation.


\autoref{fig:wqc_slice} shows the cross-sectional views for the variables $w$ and $qc$ at $1018m$ of altitude, which was selected by the domain expert. Focusing on the variable $w$, we can clearly verify that the cross-sectional views among the three members (\autoref{fig:wqc_slice} (a), (c), (e)) are completely dissimilar. This is also true for the variable $qc$ (\autoref{fig:wqc_slice} (b), (d), (f)). Considering the use of normalized values for the color mapping, and since the background color of $w$ is green in contrast to the $qc$ which is purple, thus we can understand that there is a decreasing tendency from $w$ to $qc$. By comparing the cross-sectional views of $w$ and $qc$ for the member 66 (\autoref{fig:wqc_slice} (a), (b)), we can verify a similar spatial distribution, then it is possible to infer a positive correlation between them. We can also confirm this tendency for the true state member (\autoref{fig:wqc_slice} (c), (d)), and for the member 1 (\autoref{fig:wqc_slice} (e), (f)).

\begin{figure}[tb]
    \centering
    \includegraphics[height=0.3\textheight,width=0.5\hsize,keepaspectratio]{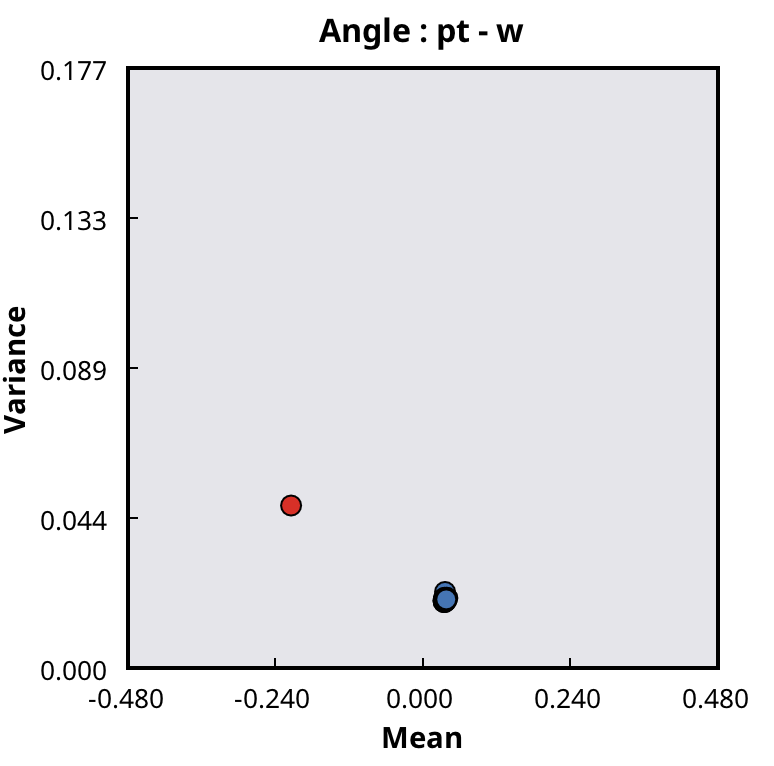}
    \caption{ADP view for the $pt-w$ interaxis region.}
    \label{fig:ptw_ADP}
\end{figure}

\begin{figure}[tb]
    \centering
    \includegraphics[height=0.6\textheight,width=1.0\hsize,keepaspectratio]{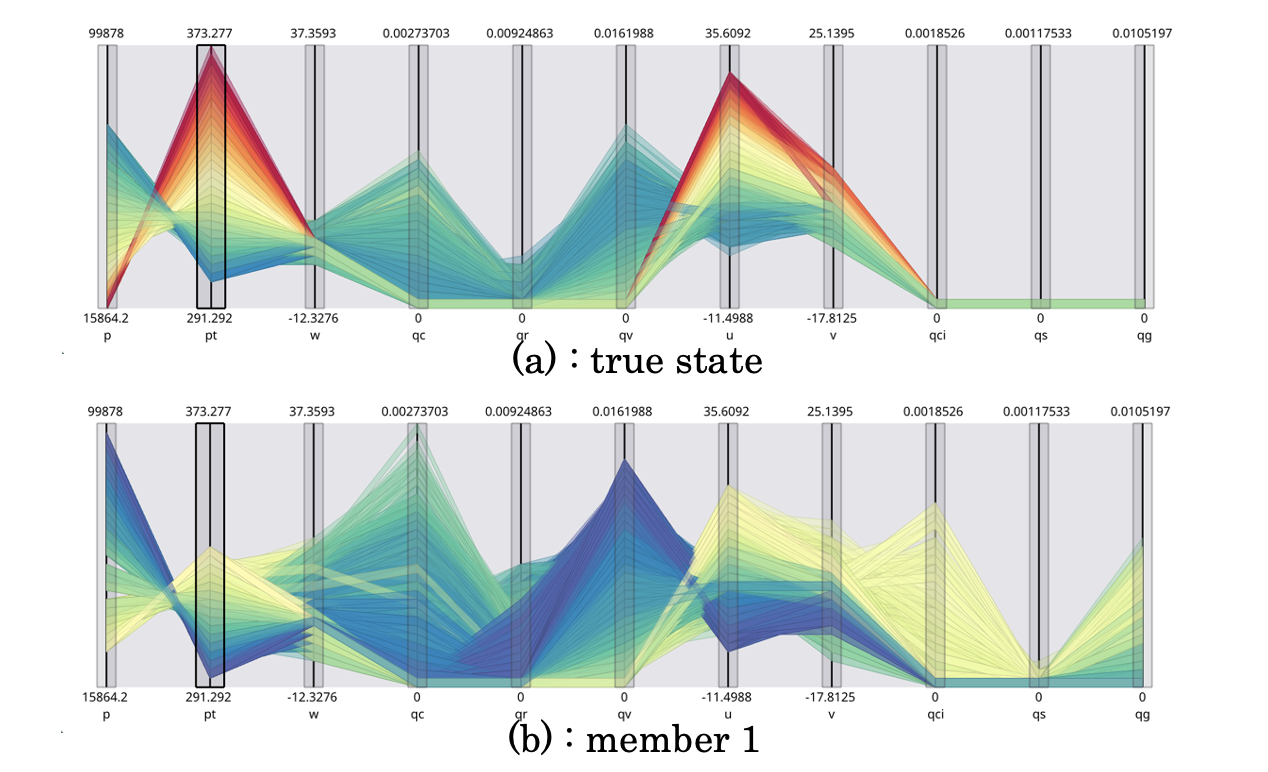}
    \caption{BPCP views focusing on the $pt$ axis. True state and member 1 are selected in (a) and (b).}
    \label{fig:ptw_BPCP}
\end{figure}

\begin{figure}[tb]
    \centering
    \includegraphics[height=0.6\textheight,width=1.0\hsize,keepaspectratio]{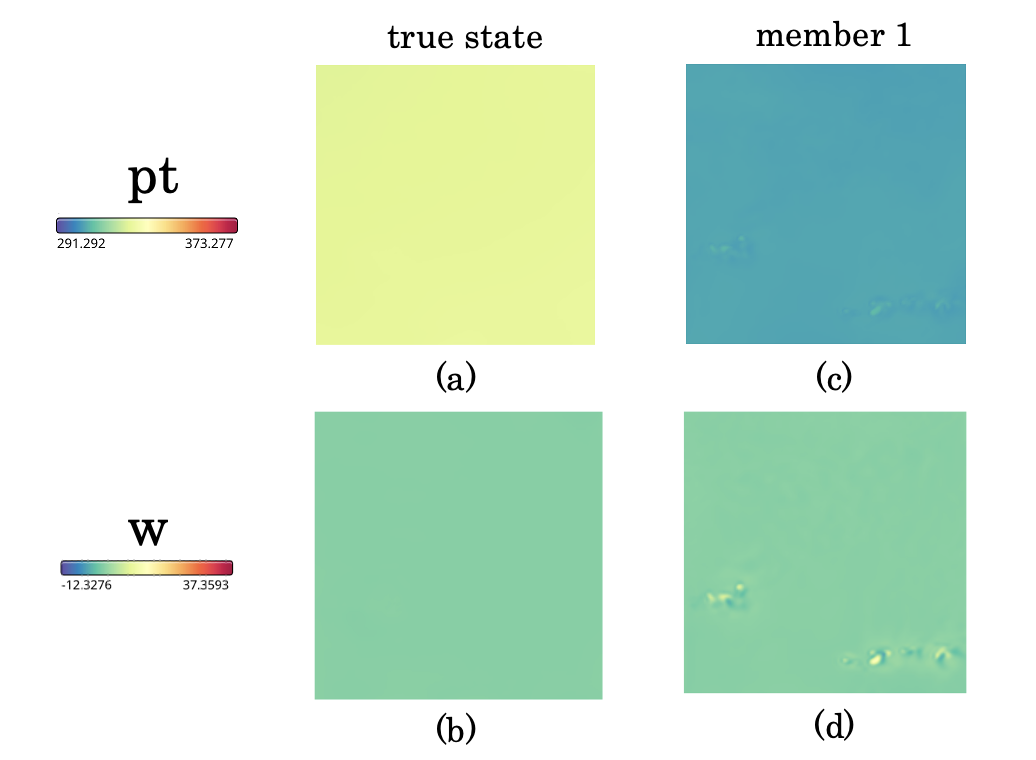}
    \caption{Cross-sectional views for the variables $pt$ and $w$ at $1018m$ of altitude.}
    \label{fig:ptw_slice}
\end{figure}

\begin{figure}[tb]
    \centering
    \includegraphics[height=0.6\textheight,width=1.0\hsize,keepaspectratio]{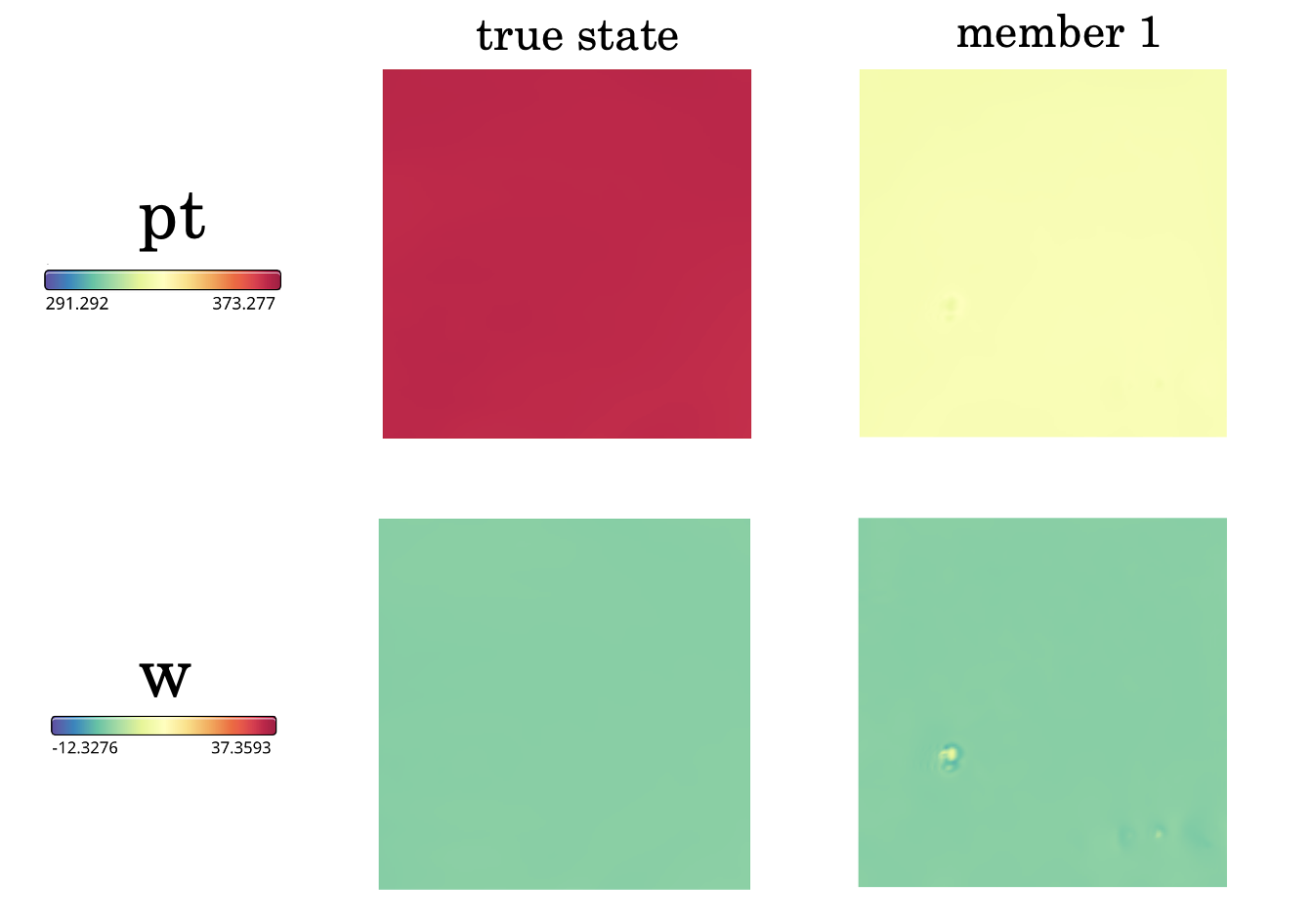}
    \caption{Cross-sectional views for the variables $pt$ and $w$ at $5506m$ of altitude.}
    \label{fig:ptw_slice2}
\end{figure}

\subsubsection{Case 2: Potential Temperature $pt$ and Vertical Component of the Wind Velocity $w$}
\label{case_2}


Focusing on the APCP and ADP results shown in~\autoref{fig:APCP_ADP} (b), and looking to the $pt$ and $w$ axes, we can verify the existence of members plotted far away from each other on both variable axes. When focusing on the ADP view in \autoref{fig:APCP_ADP} (b), we can verify that most of the scatter plots are concentrated on a single region with an exception to the one corresponding to the true state member. That is, excluding the true state member, we can presume that the crossing pattern and distributions of the line segments in the interaxis region of $pt$ and $w$ axes are similar. Looking to the scatter plot corresponding to the true state in \autoref{fig:ptw_ADP}, we can observe that the variance is larger than all other members, thus the tendency for negative correlation becomes stronger, and the mean value becomes larger in the negative direction. Similarly to the previous case study, to investigate the hypothesis, we used the BPCP view. Focusing on the BPCP results for the members 1 and the true state, shown in~ \autoref{fig:ptw_BPCP}, we can observe that the intersection pattern for the true state differs from the member 1. In addition, we can verify from the BPCP that the line segments, for the true state, are more negatively inclined in the interaxis region of $w$ and $qc$ axes than those of the member 1.




\autoref{fig:ptw_slice} shows the cross-sectional views for the variables $pt$ and $w$ at $1018m$ of altitude similarly to the previous case study. By comparing the cross-sectional views of member 1 and true state for the variable $pt$ (\autoref{fig:ptw_slice} (a), (c)), we can verify a dissimilar spatial distribution. Similarly, we can also verify that the spatial distribution is also dissimilar for the variable $w$ (\autoref{fig:ptw_slice} (b), (d)). When looking to the cross-sectional view at $5506m$ of altitude (\autoref{fig:ptw_slice2}), we can verify that for the true state the color mapping becomes red for the variable $pt$ and green for the $w$. In addition, for the member 1, it becomes yellow for the variable $p$ and green for the $w$. Therefore, we can infer that in the interaxis region of $qt$ and $w$ axes, the true state has stronger downward trend than that of member 1.

\begin{figure}[tb]
    \centering
    \includegraphics[height=0.6\textheight,width=1.0\hsize,keepaspectratio]{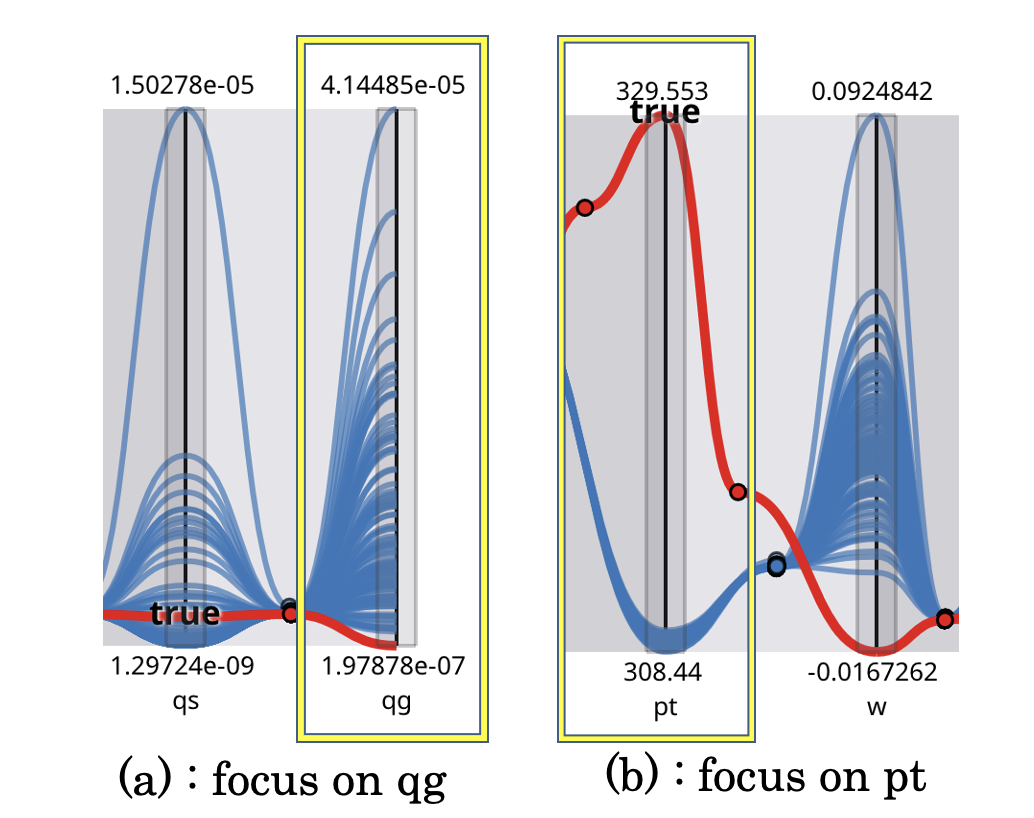}
    \caption{APCP view focusing on the variables $qg$ and $pt$.}
    \label{fig:qgpt_APCP}
\end{figure}

 As stated before, the fact of the true state having smaller value than all mean values of variable $w$, was previously noted, and at that time, the domain expert suspected that the mixing ratio of graupel ($qg$) generated this situation. The proof of this hypothesis was considered a future work, and we could use the implemented visual analytics system, with the proposed APCP, to investigate and confirm this hypothesis. The summarized results are as follows:

\begin{enumerate}
    \item The resulting ensemble forecast greatly overestimates the mean value of $qg$ compared to the true state (\autoref{fig:qgpt_APCP} (a)).
    \item The utilized data assimilation process generated excessive ice particles that excessively cooled the upper sky, thus lowering the value of $pt$.
    \item As the value of $pt$ becomes smaller, the upper sky becomes an unstable stratosphere (a situation where the convection is likely to occur). As a result, the value of $pt$ in the ensemble forecast becomes smaller than of the true state (\autoref{fig:qgpt_APCP} (b)).
    \item Since the upper sky becomes an unstable stratosphere, a large value of $w$ is predicted in the ensemble forecast.
\end{enumerate}

\vspace{5pt}


Although we could confirm the effectiveness of the proposed APCP method, there still remains a problem likely related to the PCP technique itself. The correlation between two variables can only be inferred for those variables placed as adjacent axes. Therefore, when analyzing multivariate ensemble simulation data, it may become necessary to repeatedly change the order of the variables (parallel axes) during the visual analysis. To enable a more efficient analysis, it becomes necessary to take into consideration an optimal ordering of the parallel coordinate axes. In addition, due to the limited dimensions capable to be covered simultaneously, we used pre-defined fixed time for the visual analysis. Therefore, in order to analyze in the time direction, it becomes necessary to rerun the application after each change in the time step for the analysis.

\section{Conclusion}




In this work, we proposed the Angular-based Edge Bundled Parallel Coordinates Plot (APCP) for the visual analysis of large ensemble simulation datasets. The APCP simplifies the PCP of each member into a single representative curved line connecting scatter plots possessing the angular information of the partial line segments. As a result, it becomes possible to infer the correlations between the adjacent variables while minimizing the visual cluttering problem caused by overplotting. We developed an innovative visual analytics system which implemented the proposed APCP with additional coordinated linked views for assisting the interactive visual analysis. We confirmed the effectiveness of the proposed system by executing case studies in collaboration with a domain expert. For this purpose, we used a meteorological ensemble simulation data performed by the Japanese flagship supercomputer Fugaku. Although we obtained positive feedback and encouraging results, there is a need to evaluate using much larger ensemble simulation data such as those conducted in modern meteorological simulation studies. As another future work, we will work trying to enable the use of the time information during the interactive visual analysis, that is, to enable comprehensive visual analysis of ensemble simulation data from all four facets: members, variables, space, and time.

\acknowledgments{
This work was partially supported by JSPS KAKENHI (Grant Numbers: 20H04194, 21H04903, 22H03603). This work used computational resources of supercomputer Fugaku provided by the RIKEN Center for Computational Science.}

\bibliographystyle{abbrv-doi}

\balance
\bibliography{2022_LDAV__KeitaWatanabe}
\end{document}